\documentclass[10pt]{iopart}
\usepackage{iopams}
\usepackage{hyperref}
\usepackage{fullpage}
\usepackage{graphicx}
\usepackage{color}

\usepackage{xcolor}
\definecolor{trueblue}{rgb}{0.0, 0.45, 0.81}
\definecolor{crimsonglory}{rgb}{0.75, 0.0, 0.2}

\usepackage{hyperref}
\hypersetup{colorlinks,linkcolor={trueblue},citecolor={crimsonglory},urlcolor={crimsonglory}}

\begin{document}

% Maketitle
\title[Volkov-like states in graphene]{Sound waves induce Volkov-like states, band structure and collimation effect in graphene}

\author{M Oliva-Leyva$^1$, Gerardo G Naumis$^1,^2$}

\address{$^1$ Depto. de F\'{i}sica-Qu\'{i}mica, Instituto de F\'{i}sica, Universidad Nacional Aut\'{o}noma de
M\'{e}xico (UNAM). Apdo. Postal 20-364, 01000, M\'{e}xico D.F. 01000,
Mexico}

\address{$^2$ School of Physics Astronomy and Computational Sciences, George Mason University, Fairfax, Virginia 22030, USA}

%\email[E-mail: ]{naumis@fisica.unam.mx}
\eads{\mailto{moliva@fisica.unam.mx},
\mailto{naumis@fisica.unam.mx}}
\date{\today}

\begin{abstract}
We find exact states of graphene quasiparticles under a time-dependent deformation (sound wave), whose propagation velocity is smaller
than the Fermi velocity.
To solve the corresponding effective Dirac equation,
we adapt the Volkov-like solutions for relativistic fermions in a medium under a plane electromagnetic wave.  The corresponding
electron-deformation quasiparticle spectrum is determined by the solutions of a Mathieu equation resulting in band tongues warped
in the surface of the Dirac cones. This leads to a collimation effect of electron conduction due to strain waves.
\end{abstract}

%\noindent{\it Keywords\/}: graphene, time-dependent deformation, pseudoelectromagnetic field, wave-particle interaction, Dirac equation

%\pacs{81.05.ue, 72.80.Vp, 73.63.-b}

%\keywords{graphene, time-dependent deformation, pseudoelectromagnetic field, wave-particle interaction, Dirac equation}

%\submitto{\JPCM}

\maketitle

\twocolumn

\section{Introduction}

Contrary to the parabolic dispersion of charge carriers in
most materials, the quasiparticles in graphene exhibit
a linear relation between energy and momentum, and thus
behave as massless relativistic fermions \cite{Novoselov04}.  Consequently,
the low-energy description for the quasiparticles in graphene
is given by the massless Dirac equation, with an effective ``speed of light'' of
$8\times10^{5}\mbox{m/s}$. This property results in
a number of unprecedented features, such as  Klein tunneling effect, specific integer and
fractional quantum Hall effect, a ``minimum'' conductivity of
$\sim4e^{2}/h$ even when the carrier concentration tends to zero,
weak antilocalization, high mobilities of up to $10^{6}\mbox{cm}^{2} \mbox{V}^{-1} \mbox{s}^{-1}$,
universal transmittance expressed in terms of the fine-structure constant, among others \cite{Geim09,Novoselov11,Katsnelson}.

Graphene, also shows unique mechanical properties \cite{Castellanos}. Suffice it to say that graphene has an effective Young modulus of
$\sim1\mbox{TPa}$ and simultaneously, can reversibly withstand elastic deformations up
to $25\%$ \cite{Lee08}. This unusual interval of elastic response has opened a new opportunity to explore the
strain-induced modifications of the its electrical, chemical and optical properties, and thus, to improve its
technological functionality. For example,
a band-gap opening has been achieved by using an uniaxial strain \cite{Ni08,Pereira09a}. On the other hand,
via stretching of the supporting flexible substrate, produces impressive increases in the chemical
reactivity of graphene \cite{Bissett13}. Also, the nonlinear response of nanoelectromechanical graphene resonators has opened up new device applications \cite{Isacsson10,Croy13}.
Very recently, the modulation of the transmittance for graphene under an arbitrary uniform strain
has been quantified \cite{Pereira14,Our15a}. Moreover, from a view point of basic research, strained graphene
provides a platform for studying exotic properties such as fractal spectrum \cite{Naumis14,Naumis15a}, mixed Dirac-Schr\"odinger behavior \cite{Montambaux09a,Montambaux12}, emergent gravity
\cite{Volovik,Zubkov}, topological insulator states \cite{Abanin12,Roy13}, among others \cite{Amorim}.

Nevertheless, among the most interesting strain-induced implications one can cite the experimental
observation of a spectrum resembling Landau levels in strained graphene \cite{Levy,Jiong},
which was predicted earlier by means of gauge fields \cite{Ando02,Guinea10a}. Starting from a tight-binding
elasticity approach,
in the continuum limit, one can predict the existence of these strain-induced gauge fields. Thus, a nonuniform
deformation of the lattice can be interpreted as a pseudomagnetic field, which has been deeply explored \cite{Vozmediano,Salvador13,Mucha,Zenan14,Sandler,Pereira14b,Our15b}.
However, within the same theoretical framework, a time-dependent deformation gives rise to a pseudoelectric field but its consequences have been less considered \cite{Oppen09,Firsova12,Vaezi,Sasaki14}.

The principal motivation of the present work is to
determine the consequences of a strain wave (a time-dependent deformation) on electron motion in graphene.
This paper is organized as follows. In section 2 we discuss the
effective Dirac Hamiltonian for graphene under a strain wave whereas in section 3, we obtain the
corresponding solutions. Section 4 contains discussions and conclusions.
%%%%%%%%
% = Model =
%%%%%%%%

\section{Graphene under a strain wave}

\begin{figure}
\includegraphics[width=8cm]{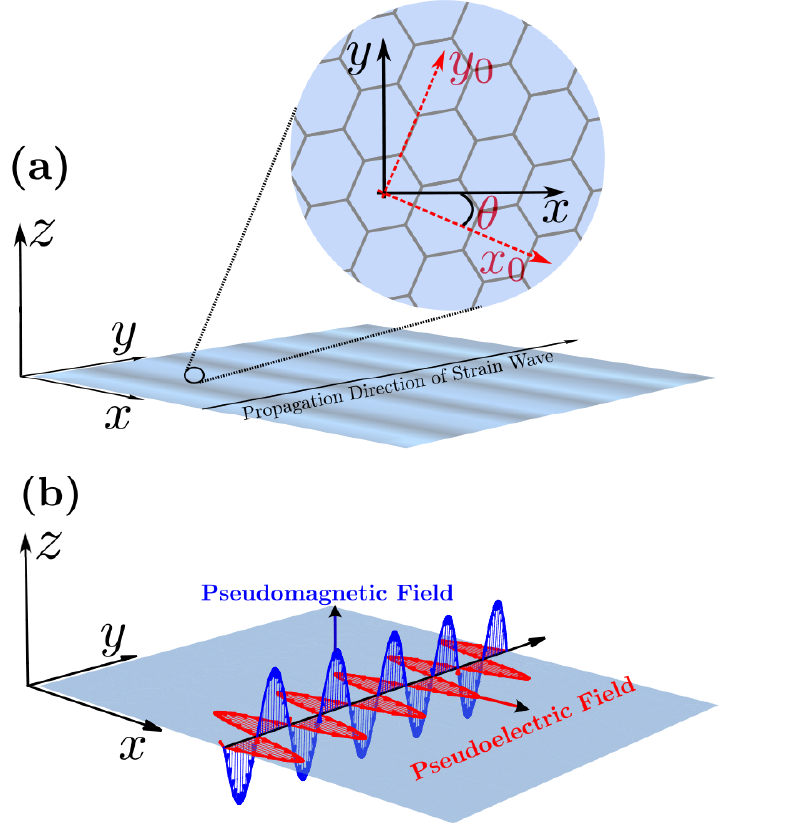} %\centerline{\includegraphics{Fig2.eps}}
\caption{\label{fig} (Color online) \textbf{(a)} Schematic representation of a sound wave in
graphene sample. The dark regions represent zones of higher density of carbon atoms
in graphene sample. In the inset, the relation between the arbitrary coordinate system $xy$ and
the crystalline coordinate system $x_{0}y_{0}$ is shown. \textbf{(b)} Unstrained graphene under a pseudoelectromagnetic wave:
an equivalent description to the problem of strain wave. The pseudoelectric field lie in graphene plane, whereas the pseudomagnetic
field is perpendicular to the graphene sample.}
\end{figure}

Consider a time-dependent deformation field of the graphene lattice $\bi{u}(x,y,t)$
described by
\begin{equation}\fl\label{def}
\bi{u}=(0,u_{0}\cos(G y- \omega t)),
\end{equation}
under the condition of continuum limit where $u_{0}\ll a_{0}\ll 2\pi/G$, i.e. the atomic displacement $u_{0}$ is much less than the unstrained carbon-carbon distance $a_{0}$ while the wavelength $2\pi/G$ is much greater than $a_{0}$.
This deformation wave propagates along the $y$ direction
with a velocity $v_{s}=\omega/G\approx2\times10^{4}\mbox{m/s}$, which is assumed to be equal to the sound velocity in graphene \cite{Adamyan}.
As illustrated in figure \ref{fig}~(a), we chose the arbitrary coordinate system $xy$ in such a way that it is rotated an arbitrary angle $\theta$ respect to
the crystalline coordinate system $x_{0}y_{0}$. For the latter, the $x_{0}$-axis points along the zigzag direction of graphene sample. Thus,
for $\theta=0$ the deformation wave moves along the armchair direction of graphene lattice, whereas for $\theta=\pi/2$, it moves along
the zigzag direction.

As discussed extensively in the literature, the electronic consequences
of a nonuniform strain can be captured by means of an effective gauge field $\bi{A}$,
which in the rotated frame $xy$ is given by \cite{Zhai},
\begin{eqnarray}\label{StandarA}
\fl A_{x}&=&\frac{\beta}{2a_{0}}\Bigl((u_{xx} - u_{yy})\cos3\theta - 2u_{xy}\sin3\theta\Bigr),\nonumber \\
\fl A_{y}&=&\frac{\beta}{2a_{0}}\Bigl(-2u_{xy}\cos3\theta - (u_{xx} - u_{yy})\sin3\theta\Bigr),
\end{eqnarray}
where $\beta\simeq3$ is the electron Gr\"{u}neisen parameter and $u_{ij}=(\partial_{i}u_{j}+\partial_{j}u_{i})/2$ is the symmetric strain tensor.
From the expression for the effective gauge field, it is clear that $\bi{A}$ exhibits a periodicity of $2\pi/3$ in $\theta$, which reflects the discrete rotational invariance,
i.e. the trigonal symmetry, of the underlying honeycomb lattice.
As is well known, in graphene there are two inequivalent Dirac cones, located at special
points of high-symmetry of the reciprocal lattice and usually denoted by $\bi{K}$ and $\bi{K}'$.
The fictitious field (\ref{StandarA}) has opposite signs at different valleys \cite{Katsnelson}. If for the valley $\bi{K}$ the effective gauge field is $\bi{A}$,
then for the other valley $\bi{K}'$ is given by $-\bi{A}$. This reflects
the fact that elastic deformations do not violate the time-reversal symmetry \cite{Morpurgo}. Here, for simplicity we will
deal with the $\bi{K}$ valley, since the other is analogous and obtainable by simply changing the sign of $\bi{A}$.

For the case of the deformation field (\ref{def}), from (\ref{StandarA}), the following effective gauge field results:
\begin{equation}\fl\label{OurA}
 \bi{A}=\frac{\beta u_{0}G}{2a_{0}}\sin(G y-\omega t)(\cos3\theta,-\sin3\theta).
\end{equation}

Hence the strain wave (\ref{def}) gives rise to a \emph{pseudoelectromagnetic wave} propagating along the $y$-axis at the velocity $v_{s}$. Here the associated pseudomagnetic field given by $\partial_{x}A_{y}-\partial_{y}A_{x}$ oscillates perpendicularly
to the graphene sample (see figure~\ref{fig}~(b)). On the other hand, the associated pseudoelectric field given by $-\partial_{t}\bi{A}$ oscillates in the sample plane
but, in general, it is not perpendicular to the propagation direction of the strain wave.
Let us illustrate this peculiar behavior.
For example, for $\theta=\pm\pi/2+2n\pi/3$, i.e. when
the strain wave moves along the zigzag direction, the pseudoelectric field oscillates along the propagation direction of the  pseudoelectromagnetic wave.
Thus, for $\theta=\pm\pi/2+2n\pi/3$, the pseudoelectromagnetic wave (\ref{OurA}) behaves more like a sort of longitudinal mechanical wave. Also,
notice that for this case the pseudomagnetic field is zero. In contrast, for $\theta=n\pi/3$, i.e. when
the strain wave moves along the armchair direction, the pseudoelectric fields oscillates transversally to the propagation direction of
the pseudoelectromagnetic wave, as the expected behavior of a real electromagnetic wave.

Then including the strain-induced gauge field via minimal coupling, the effective Dirac equation reads
\begin{equation}\fl\label{DW}
 v_{F}\bsigma\cdot(-i\hbar\nabla - \bi{A})\bPsi=i\hbar\partial_{t}\bPsi,
\end{equation}
where $v_{F}=8\times10^{5}\mbox{m/s}$ is the Fermi velocity and $\bsigma=(\sigma_{x},\sigma_{y})$
are Pauli matrices acting on sublattice space. This equation describes low-energy excitations of the electronic system
in graphene under the strain wave.

\section{Volkov-like states}

From (\ref{OurA}) one can clearly distinguish the periodicity of the effective gauge field $\bi{A}$ along the time direction.
Therefore, one could think the standard use of the Floquet theory  to analyze time-periodic Hamiltonians \cite{Hanggi}. At the same time, $\bi{A}$ presents space periodicity along the $y$-direction, so that, Bloch (Floquet)
theory could also be used. However, since $\bi{A}$ ultimately depends upon the plane phase $\phi=G y-\omega t$ of the wave,
to solve (\ref{DW}), we propose a spinor wavefunction $\bPsi$ of the form \cite{Volkov}
\begin{equation}\fl\label{PV}
 \bPsi(x,y,t)=\exp[i(k_{x}x + k_{y}y - Et/\hbar)]\bPhi(\phi),
\end{equation}
and then, a Floquet analysis could be translated to the effective equation of the spinor, as carried out in $\bPhi(\phi)$ \cite{Cronstrom}.
In relativistic mechanics, this procedure  is equivalent to a jump into the light-cone frame of reference.
In principle, $E,k_{x},k_{y}$ are just parameters of the ansatz. However,
to recover the solution for an electron in undeformed graphene, obtained as $A_0\rightarrow 0$,
it is needed that $E=\pm\hbar v_{F}(k_{x}^{2}+k_{y}^{2})^{-1/2}$, as is done in similar relativistic
problems \cite{Cronstrom,Becker,Landau}.

Our ansatz (\ref{PV}) was firstly used by Volkov to find exact solutions of the Dirac equation for relativistic fermions under a plane
electromagnetic wave in vacuum \cite{Volkov}. It is worth mentioning that only in the case when the electromagnetic wave propagates in vacuum
the solutions of the Dirac equation can be found in a simple closed form, these are the Volkov states. However, when one consider
the interaction with a plane wave propagating in a medium with an index of refraction $n_{m}\neq 1$, the mathematical complexity
of the Dirac equation is largely increased, and in fact, it is a problem with a long history
(see for example \cite{Varro13, Varro14} and references therein).

Comparing our problem, given by (\ref{OurA}) and (\ref{DW}), with the standard problem of real relativistic fermions under
a plane electromagnetic wave in a medium with an index of refraction $n_{m}$, one can recognize the following. The Fermi velocity
$v_{F}$ plays the role of the speed of light in vacuum $c$, whereas the strain wave velocity $v_{s}$ plays the role of the electromagnetic wave
speed $c/n_{m}$ in a medium.
From this analogy, one can think that our pseudoelectromagnetic wave propagates in a medium with effective index of refraction
$n_{m}=v_{F}/v_{s}>1$, which is a different limiting than the one considered in the real electromagnetic case where $n_{m}<1$ \cite{Naumis08}.
Only in the hypothetical case that $v_{F}=v_{s}$, one can obtain the usual Volkov states \cite{Varro81}.

Substituting (\ref{PV}) into (\ref{DW}) and taking into account $v_{F} \gg v_{s}$, we obtain the following
differential system for the components of spinor $\bPhi(\phi)$,
\begin{eqnarray}\label{LS}
\fl \frac{\rmd\Phi_{B}}{\rmd\phi}=(\tilde{k}_{x} - i\tilde{k}_{y} - \tilde{A}_{0}e^{i3\theta}\sin\phi)\Phi_{B} - \tilde{E}\Phi_{A},\nonumber\\
\fl-\frac{\rmd\Phi_{A}}{\rmd\phi}=(\tilde{k}_{x} + i\tilde{k}_{y} - \tilde{A}_{0}e^{-i3\theta}\sin\phi )\Phi_{A} - \tilde{E}\Phi_{B},
\end{eqnarray}
where we define the non-dimensional parameters: $\tilde{k}_{x,y}=k_{x,y}/G$,
$\tilde{A}_{0}=\beta u_{0}/(2a_{0})$ and $\tilde{E}=E/(\hbar v_{F} G)$. This system can be reduced to a
second-order differential equation for each component of spinor $\bPhi(\phi)$, denoted by ${\Gamma}_{A}$ and ${\Gamma}_{B}$. However,
before it is appropriate
to carry out a new ansatz:
\begin{equation}\fl\label{An2}
 \bPhi(\phi)=\exp[-i\tilde{k}_{y}\phi + i \tilde{A}_{0}\sin3\theta\cos\phi]\bGamma(\phi).
\end{equation}

As a result, from (\ref{LS}) and (\ref{An2}) we obtain that the both components ${\Gamma}_{A}$ and ${\Gamma}_{B}$ satisfy the
Mathieu equation (see Appendix):
\begin{equation}\fl\label{ME}
 \frac{d^{2}\Gamma_{A,B}}{d\zeta^{2}} + (a - 2 q\cos\zeta)\Gamma_{A,B}=0,
\end{equation}
where we introduce the variable
\begin{equation}\fl
 \zeta=(\phi + \phi_{0})/2,\ \ \ \mbox{with} \ \tan\phi_{0}=2\tilde{k}_{x},
\end{equation}
and the new parameters
\begin{equation}\fl\label{aq}
 a=4\tilde{k}^{2}_{y}, \ \ \mbox{and} \ \ q=2\tilde{A}_{0}(1+4\tilde{k}^{2}_{x})^{1/2}\cos3\theta.
\end{equation}

Therefore, the general solutions to the components ${\Gamma}_{A}$ and ${\Gamma}_{B}$ are linear combinations of the Mathieu cosine
$C(a,q,\zeta)$ and Mathieu sine $S(a,q,\zeta)$ functions. Nevertheless, taking into account that
when the pseudoelectromagnetic field is switched out ($\tilde{A}_{0}=0$) the wavefunction $\bPsi$ must reduce to a free-particle
wavefunction, we get
\begin{equation}\fl\label{Ga}
 \bGamma(\zeta)=N(C(a,q,\zeta) + i S(a,q,\zeta))
 \left( \begin{array}{c}
1 \\
s e^{i\alpha}
\end{array}\right),
\end{equation}
where $\alpha=\tan(\tilde{k}_{y}/\tilde{k}_{x})$, $N$ is a normalization constant and $s=\pm1$ denotes the conduction
and valence bands, respectively. To make sure that (\ref{Ga}) reproduces the case of undeformed graphene, it is enough
to consider the properties of the Mathieu functions for $q=0$: $C(a,0,\zeta)=\cos(\sqrt{a}\zeta)$ and
$S(a,0,\zeta)=\sin(\sqrt{a}\zeta)$.

\begin{figure}[t]
\includegraphics[width=7cm]{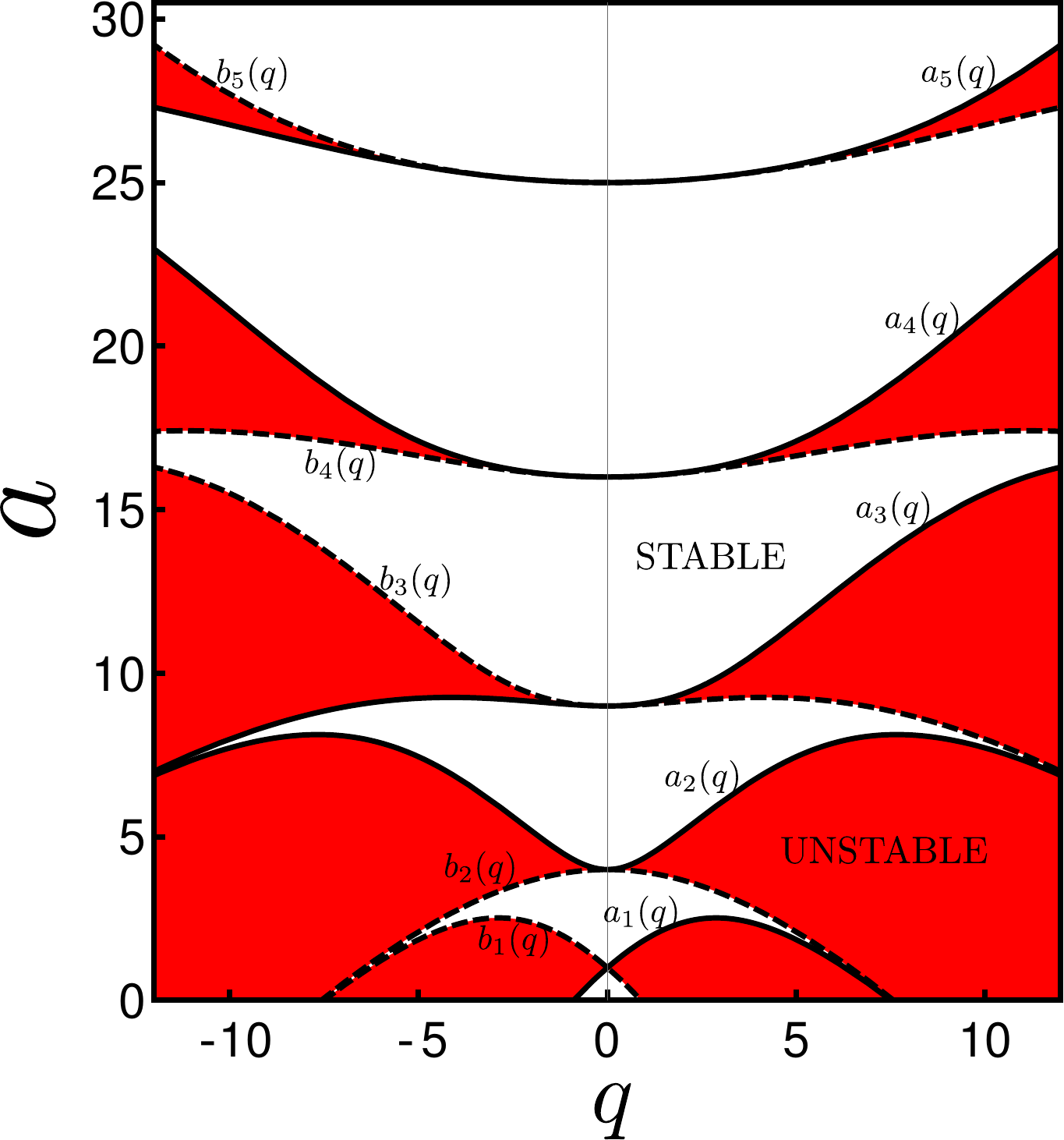} %\centerline{\includegraphics{Fig2.eps}}
\caption{\label{fig2} (Color online) Stability chart as a function of the non-dimensional parameters $a$ and $q$ for Mathieu solutions.
The regions of stability (white domains) and instability (red domains)
are divided by the characteristic curves $a_{n}(q)$ (solid lines) and $b_{n}(q)$ (dashed lines). The chart is symmetrical respect to the $a$-axis.}
\end{figure}

\section{Discussion}

As it is well documented \cite{McLachlan}, the stability of Mathieu functions depends on the parameters $a$ and $q$.
In figure~\ref{fig2}, the red regions in the ($a,q$)-plane are those for which the solutions are unstable (exponential functions)
and therefore, are not acceptable from a quantum view point. On the other hand, the white regions are those for which the solutions
are acceptable wavefunctions. The boundaries between these regions are determined by the eigenvalues, $a_{n}(q)$ and
$b_{n}(q)$, corresponding to the $2\pi$-periodic Mathieu functions of integer order, $ce_{n}(q,\zeta)$  and
$ce_{n}(q,\zeta)$, respectively \cite{McLachlan}.

\begin{figure}
\includegraphics[width=8cm]{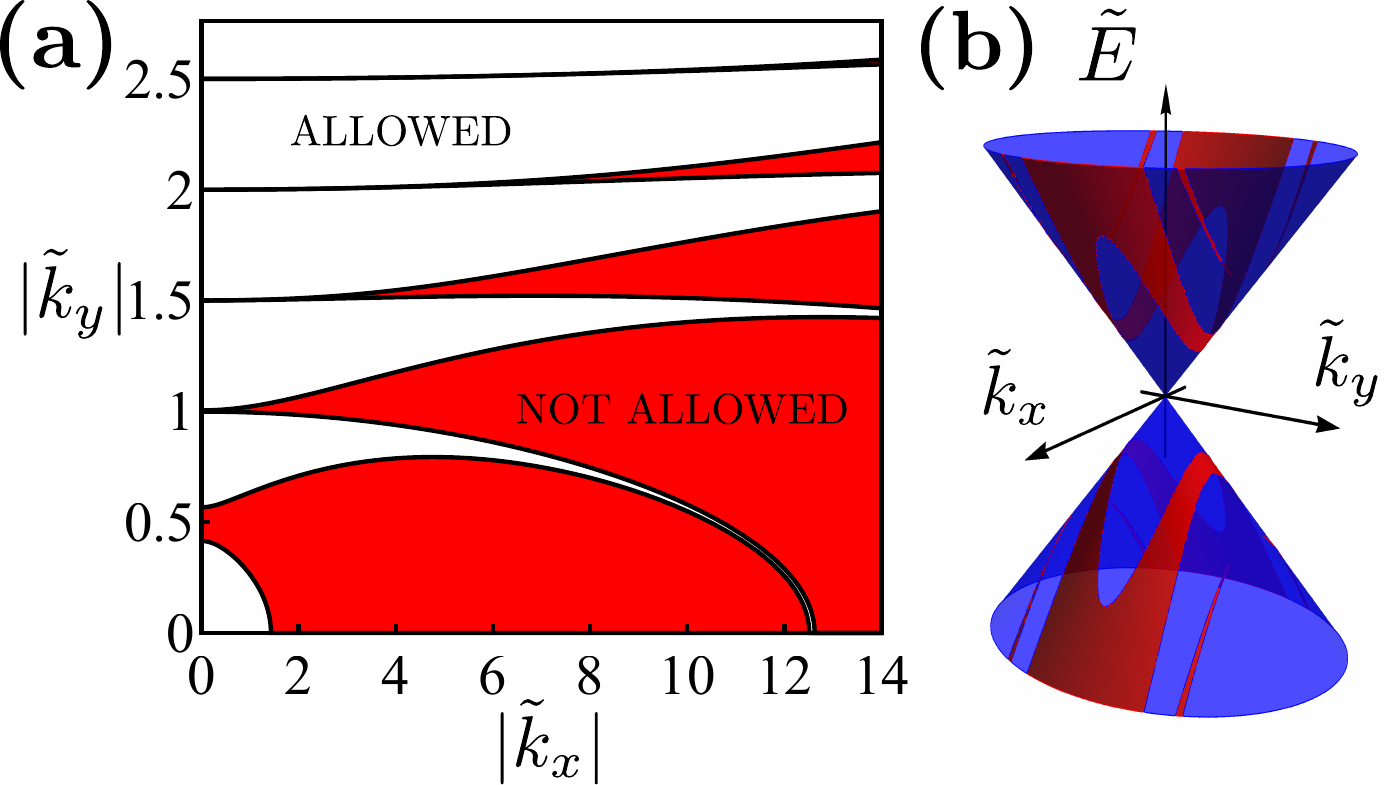} %\centerline{\includegraphics{Fig2.eps}}
\caption{\label{fig3} (Color online) Collimation effect of electron conduction by strain-waves. \textbf{(a)} Chart of the allowed bands
(white regions) for the quasi momentum $\tilde{\bi{k}}$ (in units of $G$), with $\tilde{A}_{0}=0.15$
and the strain wave propagating along the armchair direction. The diagram is symmetrical respect to both axes. \textbf{(b)} Dirac cone. Red (blue) strips correspond to the forbidden
(allowed) values of the
quasi-energy $\tilde{E}(\tilde{k}_{x},\tilde{k}_{y})$ (in units of $\hbar v_{F}G$), because of the strain wave.}
\end{figure}

As a consequence from the properties of Mathieu solutions mentioned above, a ``band structure'' naturally emerges in our problem.
If one take into account (\ref{aq}), the stability chart in the ($a,q$)-plane is translated into a chart of the allowed
bands in the ($\tilde{k}_{x},\tilde{k}_{y}$)-plane, as shown in figure~\ref{fig3} (a). When the strain wave
propagates along the zigzag, $\cos3\theta=0$, then $q=0$ and therefore the quasi-wave vector $\tilde{\bi{k}}=(\tilde{k}_{x},\tilde{k}_{y})$
can take any value. However, from the dependence $q\sim\cos3\theta$, one can conclude
that for a strain wave propagating along the armchair direction ($\cos3\theta=\pm1$), the band gaps in the
($\tilde{k}_{x},\tilde{k}_{y}$)-plane are expanded. Just for the sake of illustration, hereafter we consider $\theta=n\pi/3$.

In figure~\ref{fig3}~(a), we display the allowed values of the quasi-wave vector
$\tilde{\bi{k}}=(\tilde{k}_{x},\tilde{k}_{y})$ for a strain wave of amplitude $u_{0}=0.1a_{0}$ and moves along the armchair direction.
The most important result to notice
in figure~\ref{fig3}~(a) is that the quasi-particles propagate preferably in the $y$ direction, i.e.,
in the propagation direction of the strain wave leading to a collimation effect of the electrons. For example,
notice that if $\tilde{k}_{y}=0$, the allowed values of $\tilde{k}_{x}$ are
practically limited to the interval ($-q_{c}/4A_{0},q_{c}/4A_{0}$), where $b_{1}(q_{c}\approx0.91)=0$. In other words, only low energy
quasi-particles propagate perpendicularly to the propagation direction of the strain wave.
On the contrary, for $\tilde{k}_{x}=0$,  $|\tilde{k}_{y}|$ can take all values except basically those of
the form $m/2$, where $m$ is a positive integer.  This last result is an analogous condition as Bragg's
diffraction.

To end, let us point out that the forbidden values of the quasi-wave vector $\tilde{\bi{k}}$  divide the Dirac cone,
given by $\tilde{E}=\pm(\tilde{k}_{x}+\tilde{k}_{y})^{1/2}$, into strips of allowed and forbidden values of the quasi-energy $\tilde{E}$,
as illustrated in figure~\ref{fig3}~(b).
Similar results for the band structure have been discussed in earlies works for the case of a spin-less
particle (which obeys the Klein-Gordon equation) in a medium (of $n_{m}>1$) irradiated with an electromagnetic plane wave \cite{Cronstrom,Becker}.
On the other hand, our results differ from those
reported for graphene under an electromagnetic plane wave moves with the velocity of light in vacuum $c\approx300 v_{F}$ \cite{Naumis08}. The physical
reason is simple. In our problem, the pseudoelectromagnetic wave moves with the velocity of sound $v_{s}\approx v_{F}/40$.

Vaezi \emph{et al} \cite{Vaezi} reported the possibility of observing a nonvanishing charge current in graphene by applying a time-dependent strain.
To archive this edge charge current, it was assumed a mass term to provide the presence of a gap in the spectrum, which
is an essential ingredient to get a quantized response. From figure~\ref{fig3}~(b) one can be distinguished that graphene remains gapless.
Therefore, it does not seem possible to archive such edge charge currents in graphene under the considered approximations. To achieve
a gap and thus topological modes as happens in the electromagnetic case \cite{Naumis08}, this will require a much higher amplitude of the strain than the one
considered here since intervalley mixing is needed.

This is a consequence that in fact, there
is a crucial difference between magnetic and pseudomagnetic
fields. As commented above, the coupling constant for  pseudomagnetic fields
has opposite signs for electrons at different valleys, whereas for a magnetic
field is the same. Since the coupling constant is valley anti-symmetric, the currents
at the $\bi{K}$ and $\bi{K}’$ valleys flow in the opposite directions and they cancel
out \cite{Sasaki14}. As a result, in general a pseudoelectric field does
not cause a net electric current and thus adiabatic pumping is not possible as happens
in the real electromagnetic fields \cite{Naumis08}. Notice that in our case, the pseudomagnetic
field can be considered as adiabatic since $\omega\ll G v_F$.

\section{Conclusion}

In conclusion, we studied the effects of a strain wave on electron motion in graphene. The coupling between the quasi-particles and the
strain wave
are captured by means of an effective pseudoelectromagnetic wave. As solutions to the resulting effective Dirac equation, we found
Volkov-type
states, which propagates preferably in the propagation direction of deformation. Also, we reported a band structure of allowed and not
allowed
values for the quasi-momentum and for the quasi-energy. The form of the emergent band structure depends on the propagation direction
of the strain wave respect to the crystalline directions of graphene lattice. This fact produces a collimation effect of charge carriers by strain waves,
which should be an alternative mechanism to archive electron beam collimation, beyond magnetic focusing \cite{Houten}, an external periodic
potential \cite{Louie08} or nanostructured heterodimensional graphene junctions \cite{Wang10}.

%%%%%%%%
% = Acknowledments =
%%%%%%%%
\ack {This work was supported by UNAM-DGAPA-PAPIIT, project IN-$102513$. M.O.L
acknowledges support from CONACYT (Mexico). G. Naumis thanks a DGAPA-PASPA scholarship for a
sabbatical leave at the George Mason University.}

%%%%%%%%%%%%%%%%
% = Appendix
%%%%%%%%%%%%%%

\appendix

\section*{Appendix}

In this section, we present the details of the calculations to derive (\ref{ME}) from (\ref{LS}).
Note that, the differential system (\ref{LS}) can be rewritten as
\begin{eqnarray}\label{LS_A}\fl
\frac{d\Phi_{B}}{d\phi}=(\tilde{k}_{x} - \tilde{A}_{0}\cos3\theta\sin\phi - i\tilde{k}_{y} - i\tilde{A}_{0}\sin3\theta\sin\phi)\Phi_{B}\nonumber\\
\fl\qquad - \tilde{E}\Phi_{A},\nonumber\\
\fl-\frac{d\Phi_{A}}{d\phi}=(\tilde{k}_{x} - \tilde{A}_{0}\cos3\theta\sin\phi + i\tilde{k}_{y} + i\tilde{A}_{0}\sin3\theta\sin\phi)\Phi_{A} \nonumber\\
\fl\qquad - \tilde{E}\Phi_{B}.\nonumber
\end{eqnarray}

To simplify this system, one can propose that
\begin{eqnarray}\label{An2_A}\fl
\bPhi(\phi)=\exp[\int^{\phi}(-i\tilde{k}_{y} - iA_{0}\sin3\theta\sin\phi^{*})d\phi^{*}]\bGamma(\phi),\nonumber\\
\fl\qquad =\exp[-i\tilde{k}_{y}\phi + i A_{0}\sin3\theta\cos\phi]\bGamma(\phi),\nonumber
\end{eqnarray}
 and then one get the following differential system
 \begin{eqnarray}\label{LSG_A}
\fl\frac{d\Gamma_{B}}{d\phi}=(\tilde{k}_{x} - \tilde{A}_{0}\cos3\theta\sin\phi)\Gamma_{B} - \tilde{E}\Gamma_{A},\nonumber\\
\fl-\frac{d\Gamma_{A}}{d\phi}=(\tilde{k}_{x} - \tilde{A}_{0}\cos3\theta\sin\phi)\Gamma_{A} - \tilde{E}\Gamma_{B},\nonumber
\end{eqnarray}
for the components of the spinor $\bGamma(\phi)$. Now, taking second derivate one can reduce this last system to
a Hill differential equation for both components $\Gamma_{A}$ and $\Gamma_{B}$:
\begin{equation*}
 \fl\frac{d^{2}\Gamma_{A,B}}{d\phi^{2}} + [\tilde{E}^{2}-\tilde{A}_{0}\cos3\theta\cos\phi - (\tilde{k}_{x} -
 \tilde{A}_{0}\cos3\theta\sin\phi)^{2}]\Gamma_{A,B}=0.
\end{equation*}

However, since in the our problem $\tilde{A}_{0}=\beta u_{0}/(2a_{0})\ll1$, the second-order terms in $\tilde{A}_{0}$ can be neglected,
and taking into account that $\tilde{k}_{y}^{2}=\tilde{E}^{2} - \tilde{k}_{x}^{2}$, then one find that
\begin{equation*}
 \fl\frac{d^{2}\Gamma_{A,B}}{d\phi^{2}} + [\tilde{k}_{y}^{2} -
 \tilde{A}_{0}\sqrt{1+4\tilde{k}^{2}_{x}}\cos3\theta\cos(\phi+\phi_{0})]\Gamma_{A,B}=0.
\end{equation*}
where $\tan\phi_{0}=2\tilde{k}_{x}$. Finally, if one carry out the variable change $\zeta=(\phi + \phi_{0})/2$ and define the parameters
$a=4\tilde{k}^{2}_{y}$ and $q=2\tilde{A}_{0}(1+4\tilde{k}^{2}_{x})^{1/2}\cos3\theta$, one obtain (\ref{ME}), which is the Mathieu equation.

%%%%%%%%
% = Bibliography =
%%%%%%%%
\section*{References}

\bibliographystyle{unsrt}
\bibliography{biblioStrainedGraphene}

\end{document}